\documentclass{aa}

\usepackage{epsfig}
\usepackage{calc}

\def\eta{et al. }

\begin{document}
\thesaurus{11.03.1, 11.06.2, 12.04.1}

\title{The fundamental plane of clusters of galaxies:\\
a quest for understanding cluster dynamics and morphology}
\author{Christoph Fritsch\inst{1} \and Thomas Buchert\inst{2,3}}
\offprints{T. Buchert; email: buchert@stat.physik.uni-muenchen.de}
\institute{Max--Planck--Institut f\"ur Extraterrestrische Physik,
Giessenbachstrasse, D--85740 Garching, Germany
\and
Theoretische Physik, Ludwig--Maximilians--Universit\"at,
Theresienstr. 37, D--80333 M\"unchen, Germany
\and
Theory Division, CERN, CH--1211 Geneva 23, Switzerland}

\date{Received 16 June 1998 / Accepted 3 December 1998}

\titlerunning{The fundamental plane of clusters of galaxies}

\maketitle

\begin{abstract}

We discuss implications of the fundamental plane
parameters of clusters of galaxies derived from combined
optical and X--ray data of a sample of $78$ nearby clusters.
In particular, we investigate the dependence of these parameters
on the dynamical state of the cluster. We introduce a new
concept of allocation of the fundamental plane of clusters derived from
their intrinsic morphological properties, and put some theoretical
implications of the existence of a fundamental plane into perspective.

\end{abstract}

\section{The fundamental plane on galaxy scales}

The concept of the {\it fundamental plane} has its origin in some
monovariant relations between global observables in stellar systems.
The Tully--Fisher relation for spirals and the
Faber--Jackson relation for ellipticals reveal that the
orbital velocity increases strongly with the luminosity.
However, the residual scatter suggests the introduction
of  an additional observable, the effective radius, which finally leads
to correlations in a three dimensional parameter space of observables,
resulting in the concept of the {\it fundamental plane}
(Dressler et al., 1987, Djorgovski \& Davis 1987).
Subsequently, this concept has been successfully employed to assess the
physical state of an elliptical galaxy.
Observationally this plane is well
established in the optical energy range (e.g. Guzman et al. 1993)
and the infrared energy range (e.g. Pahre et al. 1995) separately,
and is defined by a multivariate
correlation between characteristic parameters of elliptical galaxies.
Others like spiral galaxies have been shown to populate the same plane
reaching down to the dwarfs by suitably adopting the parameter space
(e.g. Jablonka et al. 1996).

Beyond the claim of existence of such a plane in some (observational,
but not necessarily physical) parameter space, it has proved to be a powerful
concept to also investigate the physical properties as well as the
evolutionary state of the galaxy (e.g. Bahcall et al., 1995).
Moreover, it can be employed as a
distance indicator (e.g. Van Albada et. al., 1995).
Thus, the {\it fundamental plane} has attained the status of a
physical concept, i.e. a (conjectured) relation among physical variables.

\section{Reaching out to the realm of cosmology}

It is a natural question to ask whether such a concept may be extended to
larger spatial scales, but there are also strong implications of such an
extrapolation which have to be seriously considered.

\subsection{The fundamental plane on cluster scales}

Although clusters of galaxies mainly consist of
elliptical galaxies, this is not a straightforward argument to expect
that such a plane can be found for structures on cluster scales.
Schaeffer et al. (1993) have advanced that claim, however, based on
a small set of $25$ optically selected clusters. Recently,
Adami et al. (1998) have consolidated this concept
with a sample of $29$ clusters within the ESO nearby Abell cluster survey.
In the present note we advocate this claim.
However, we would like to stress that the luminous matter in clusters of
galaxies is mainly built from the X--ray emitting intergalactic gas
(10\% - 30\%) and not from galaxies (1\%-10\%);
the allocation of a {\it fundamental plane}
should therefore be decided on the basis of X--ray data rather than optical data
alone.
Their combination carries additional information about the coupling between galaxies
and intergalactic gas, which are both supposed to be sett\-led down
in the gravitational potential built from the hitherto unspecified `dark matter'
(70\%-90\%).
Based on the work by Fritsch (1997) and Fritsch \& B\"ohringer (in prep.), 
which supports the existence of a {\it fundamental plane}
in the characteristic optical and X--ray properties of galaxy clusters, we
are going to investigate implications of morphometric parameters
that quantify cluster substructure.

There is also a dynamical reasoning for the existence of a fundamental
plane which assumes the scalar virial theorem, which was derived for isolated systems,
to hold. This idealization provides a relation between the
relevant physical parameters, here: the gravitational mass,
the extent of the structure under consideration
(measured in terms of half--light radius, see below) and the stabilizing dynamical
pressure due to velocity dispersion. We shall also consider this relation
below to derive values for the mass--to--light ratio.
Although simple--minded, the use of this relation stems from the common
implication that an ensemble of ``virialized'' entities should define the
{\it fundamental plane}.
\noindent
Can this be so simple?

\subsection{Theoretical implications}

While galaxies are easily identified as individual entities, although their
dark halos may not fully follow that identification, the cluster as a structural
unit is more difficult to assess.
People tend to consider clusters of galaxies as the largest gravitationally
bound systems in the Universe, and the wordings ``decoupled from the universal
expansion'' and ``relaxed, or virialized system'' are applied to suggestively
establish the possibility of isolation from the
large--scale structure environment. It is clear that this can only be true
for an idealized cluster; real ones are neither ``relaxed'', nor ``decoupled'' from
the expansion and their environment.
To write down the simplest possible formula for the relation between
physical variables (the scalar virial theorem), and to use this relation as
a standard of reference of a ``relaxed'' cluster is,
at least, courageous.

\smallskip\noindent
{\it What is a cluster?}

\noindent
In general we may consider any overdense patch of matter and require some
stationarity condition to hold for it {\it on average}. This defines
the term ``decoupling from the expansion''.
The tensor virial theorem as defined by Chandrasekhar \& Lee (1968) cannot be applied
{\it per se} to a non--isolated system. Instead a generalized
cosmic virial theorem should state the
relation between spatially averaged quantities for some spatial domain
{\it embedded} into the cosmological model. This, still,
requires the treatment of boundary terms that arise by averaging over an
overdense portion of matter at the boundaries of the averaging domain.
Within the cluster all dynamical
variables in general fluctuate, and in turn the strength of the fluctuations
themselves has influence on the average properties. The latter is known as the
``backreaction effect'' (see, e.g., Buchert \& Ehlers 1997 and ref. therein).
The picture to be emphasized here is that a cluster
which is {\it embedded} into the large--scale structure environment
should be subjected to a stationarity condition on average taking the
internal dynamics {\it and} cluster boundary terms into account (compare here the
so--called C--correction that has been recently applied by Girardi et al. 1998).

\smallskip\noindent
{\it The merging problem:}

\noindent
Dynamically, since we face the problem of merging, we may pick that specific
clump of matter belonging to a stationary system and trace this matter back
into the past by conserving its mass (the Lagrangian point of view). We so
avoid that there is matter in-- or outflow accross the clusters' boundary.
The evolution of morphological properties of this clump of matter may then
be considered, mapping the dynamical state as a function of time
in terms of morphological parameters.
It is here, where the idea of using morphological descriptors may provide a
landmark of how to diagnose the parameter space in which we want to allocate
a {\it fundamental plane}.

\smallskip\noindent
{\it The morphology--cosmology connection:}

\noindent
The study of the evolution of cluster morphology is a lively debated subject
in connection with cosmological simulations. Especially the relation to the
background cosmology is the focus of interest in this field (Evrard et al. 1993,
Mohr et al. 1995, Crone et al. 1996).
Still, we consider it important to understand the
notion of a ``relaxed'' cluster from first principles.
If we talk about the existence of a {\it fundamental plane}, then we imply this only
for some asymptotic dynamical state of ``relaxed'' clusters for
which the relation
implied by the {\it fundamental plane} is exactly satisfied. This relation must
be sought theoretically.
We, here, see an intimate link between the dynamics of the cluster and
a conjectured attractor in the space of the characterizing averaged variables.

We suggest to quantify the deviation from these yet unknown
defining properties of ``relaxed'' clusters by
instrinsic morphological properties such as their amount of substructure.
We explain this idea now for a sample of
clusters based on optical and X--ray data.

\section{The optically selected and X--ray based cluster data}

\subsection{The sample}

Using the COSMOS--/APM-- and the ROSAT X--ray data
robust optical and X--ray parameters for a subset of $78$ clusters of galaxies
from surface brightness profiles were derived in (Fritsch 1997; 
for details concerning the data see Fritsch \& B\"ohringer, in prep.).
To avoid the influence of evolutionary effects a homogeneous
sample of clusters of galaxies in a redshift range of $0.02 \le z \le 0.05$
was studied. A family of parameters was so derived
that, at first phenomenologically, characterize the physical state
of the clusters.
The most important independent parameters are the luminosities and the
half--light radii $R_0$ (optical) and $R_x$ (X--ray).

Fitting ellipses to each of the projected distributions gives us
the optical/ X--ray centers of the clusters. The calculation of the background is based
on the local galaxy/ photon distribution outside the cluster.
Background corrected, differential surface brightness profiles
allow us to measure the radii that contain the total light of the clusters.
Finally, the half light radii were calculated from integrated surface brightness
profiles. They define the radii within which half the light of the cluster is emitted.
The total light $L_0$ of the clusters emitted from the galaxies within
the optical blue band is given by all the background corrected galaxy magnitudes
within a circle that contains all the light.
To calculate the light below the given magnitude limit we apply the Schechter luminosity function with $M_{\star}$ = - 21.8 and $\alpha$ = -1.25.

The total X--ray luminosity was calculated iteratively by using the
Raymond--Smith code (Raymond \& Smith 1977) for a completely ionized plasma with a
typical metallicity of half the solar one, and
the empirical correlation between the X--ray luminosity and the temperature
$kT \propto L_{x}^{0.354}$ given by White (1996).
The latter correlation was used because ROSAT does not provide 
a reasonable determination of the temperature and the literature 
offers too few X--ray temperatures for a correlation analysis 
as far as our clusters are concerned.  
Superpositions of other sources were detected via
cross correlation with clusters stored in the NED--database
(NASA/IPAC Extragalactic Data Base) and cut out, subsequently.

For all the uncertainties of the derived parameters
the errors due to spatial binning and the errors of the source
and background count statistics were included.

The complete list of the 79 clusters is stored in Table~\ref{sample}.

\subsection{Morphological method of allocation}

That clusters of galaxies are not
arbitrarily distributed in the three--dimensional parameter space
$\lbrace L_0, L_x, R_0 \rbrace$ is not
surprising, but that they lie within a fairly well--defined plane is
enough reason to introduce the concept of the
{\it fundamental plane of clusters}.
To find out the physical meaning of this phenomenological ``fundamental plane''
is another issue. As outlined above we may approach the problem from first
principles. Here, we would like to sketch a procedure which already lays down a
fairly unique way to establish a diagnostic criterion of allocating the
{\it fundamental plane}. We stress, however, that in practice the sample of
clusters has to be larger than the one we are going to study in order to get
statistically significant results.

Already from visual inspection we appreciate that
the clusters can be classified in terms of some intrinsic
structural property: the clusters should pass the test whether they are
useful to define the {\it fundamental plane} in a physical sense.
We have to find a measure of the dynamical state of the
clusters of galaxies. Considering the process of structure
formation and the relaxation process for the galaxies and the gas,
a dynamical state may be represented by some morphological measure which
characterizes the amount of substructure in the clusters.
We adopt the working hypothesis that ``relaxed'' clusters of galaxies
show a small amount of substructure and
``unrelaxed'' ones, which are still in the process of merging, show
strong substructure.
Useful substructure measures have already been proposed and employed in the
literature (e.g., Crone \eta 1996 and ref. therein). We, here, base
our measure on radial variations of morphological parameters derived from
fitting ellipses to the
projected distribution of the galaxies and X--ray photons which includes
the center--of--mass shift, the ellipticity and the position angle
of cluster contours (Fritsch 1997).
In this line, robust structure functions based on vector--valued Minkowski
functionals have been proposed recently (Beisbart \& Buchert 1998) and are
currently tested on simulated clusters. This new method will also help
to overcome possible biases in the presently used morphological method 
which does not distinguish substructure from ``twisted'' isocontours.

\subsection{Laying down the {\it fundamental plane} of clusters}

Taking these substructure measures we can divide
the whole sample of clusters due to the amount
of their substructure into two classes with the same number of members.
Now we consider the polynomial $P_{EP}$ approximating the data for $L_0$:
$P_{EP}(R_0 ,L_x) = R_{0}^{0.84} \; L_x^{0.21}$ (Fig. 1). It
results from a two--dimensional $ \chi ^{2} $-fit according to the Levenberg--Marquart
algorithm to all the cluster data in the
three--dimensional parameter space, consisting of the optical luminosity $L_0$,
the X--ray luminosity $L_x$ and the optical half--light--radii $R_0$
(we call this the {\it empirical plane of clusters}). The orthogonal
scatter of that plane is given by 24 \%.
Applying a correlation analysis for both classes separately
between the optical luminosity $L_0$ and the
function $P_{EP}(R_{0},L_x)$
we determine the probabilities for the null hypothesis
that there is no correlation between $P_{EP}(R_{0},L_x)$ and $L_0$; we obtain
P$(\tau_{P_{EP};}$$L_0) \sim 10^{-6}$ for the clusters with less
substructure, and P$(\tau_{P_{EP};L}) \sim 10^{-3}$ for the clusters with much
substructure\footnote{$\tau_{P_{EP};L_0}$ is Kendall's $\tau$
quantifying the correlation between $P_{EP}(R_{0},L_x)$ and $L_0$ in a
non--parametric way.}.
Therefore, the former class may serve as that
sample which phenomenologically comprises the ``relaxed'' clusters
(defining the {\it fundamental plane}) with a reduced orthogonal scatter
of 21 \%.
Additionally we find a strong correlation between the
distance of the clusters to the so--defined {\it fundamental plane}
and the corresponding substructure measure (Figs. 1,2).
This correlation supports the
hypothesis that the location of the {\it fundamental plane} is
related to the feature of `less substructure'. \\
Whether this morphological
property uniquely relates to the virial condition, and whether the
location of the {\it fundamental plane} is reflected correctly by the
standard use of the virial relations among the parameters is not
clear at all (see also Adami et al. 1998). However,
in order to infer information on the mass--to--light ratio, we have to apply
the virial condition in its usual, albeit naive form: to estimate the masses
of the clusters of galaxies we use
the empirical relations between velocity dispersion and temperature in
consistency with the thermodynamical equilibrium condition ($\sigma^2 \propto T$),
and between temperature and X--ray luminosity ($T \propto L_x^{0.354}$)
given by White (1996).
Additionally, we have to assume a density profile to
find the relation between the optical half--light radius
and the gravitational radius $R_g$.
For simplicity we model the distribution of the
whole matter with a King profile; note, however, that the cluster mass may
be over--/underestimated by a factor of a few,
if the true density profile is steeper/shallower
in the core of the cluster (see: Sadat 1997).
Then we start from the observed correlations
between the parameters and our morphology--based allocation of the
plane and transform these parameters into the physical parameter space
${M,\sigma,R_0}$ (Fig.~3).

\subsection{Mass--to--light ratios for clusters of different morphology}

\noindent
Given the observed relations and the standard virial
condition, $\sigma^2 = \frac{GM}{R_g}$,
the {\it empirical plane} based on the total sample can be represented
by the characteristic mass to optical light relation:
\begin{eqnarray}
\frac{M}{L_0} &=&  a  \left(\frac{M}{10^{15} M_{\odot}}\right)^{b} \;\;
\left(\frac{L_x}{10^{44} {\rm erg s}^{-1}}\right)^{c}
\left(\frac{M_{\odot}}{L_{\odot}}\right)\;\;.\nonumber
\end{eqnarray}
with coefficients listed in Table~\ref{coeff} under `EP'.

From Table~\ref{coeff} we infer that there is a clear trend of discrimination
between
clusters with much substructure (index `SP') and
clusters which belong to the class of more ``relaxed'' clusters
(i.e. with less substructure in the optical and X--ray energy range)
(index `FP'). However, the respective data sets still overlap
within the errors which is a consequence of the very small sample of
clusters on which we base our analysis.
Even if we don't admit any significant difference between the {\it empirical}
and {\it fundamental} planes, there is a trend that the characteristic
mass--to--light
ratio for the ``relaxed'' clusters depends only slightly on the mass
and almost does not depend on X--ray luminosity;
it suggests a constant value for $M/L$ which is a striking result of
our morphology--based allocation of the {\it fundamental plane}. The latter
would also support the
assumption that the light distribution of the galaxies follows the
mass distribution in ``relaxed'' clusters, an assumption that lies on
the basis of most mass estimates (see, e.g., Girardi et al. 1998).
Furthermore, in comparison with the mass--to--light ratio for
elliptical galaxies (Bender et al. 1992), our data imply that
the mass--to--light ratio of clusters reveals about 30 times more ``non--optical
luminous mass'' than the elliptical galaxies which gives a hint to a certain
fraction of the X--ray mass and especially to the amount of the underlying
unknown dark matter (Fig.~4).  It is interesting that our analysis suggests a tendency
towards lower values for $M/L$ in clusters as we approach the
{\it fundamental plane}.

\begin{table}

\caption{Coefficients for the mass--to--light ratios
for the subsamples `EP' ({\it empirical plane}, all clusters), `FP'
({\it fundamental plane}), `SP' (clusters with much substructure).}

\begin{minipage}[b]{.99\linewidth}
\centering
\begin{tabular}{|c|c|c|c|}
\hline
index & $a$ & $b$ & $c$\\ \hline \hline
  EP  & $448.70 \pm 148.21$ & $ 0.16 \pm 0.17$ & $0.09 \pm 0.12$\\ \hline
  SP  & $613.16 \pm 190.17$ & $ 0.30 \pm 0.14$ & $0.08 \pm 0.15$\\ \hline
  FP  & $318.32 \pm 107.13$ & $ 0.12 \pm 0.18$ & $0.10 \pm 0.13$\\ \hline
\end{tabular}

\end{minipage}\hfill

\label{coeff}

\end{table}

\section{Conclusions}

The data show that the nearby ($0.02 \le z \le 0.05$)
clusters of galaxies lie preferentially within a  plane in a three--dimensional
parameter space built from the optical luminosity, the X--ray luminosity
and the optical half--light radius. On the basis of a morphological criterion
for cluster substructure and the hypothesis that ``relaxed'' clusters reveal less
substructure, we identified a {\it fundamental plane} by a 2--dimensional fit to a (by a factor of
1/2) reduced sample of clusters with less substructure.
The distances of the other clusters to this {\it fundamental plane} show strong correlations
with our measure of substructure. The proposed method for allocating the
{\it fundamental plane}
should be viewed in parallel with a physical condition among spatially averaged dynamical
variables, which still has to be sought. Adopting the generally held view of
the standard virial relations (based on the trace of the tensor virial theorem for
isolated systems), we derived the $M/L$ values for each class of clusters implying a
tendency towards lower values for more ``relaxed'' clusters. Our method
suggests a weak dependence on mass and luminosity for clusters populating the
{\it fundamental plane} resulting in an almost constant value for $M/L$
of typically $300$.
The (orthogonal) scatter around the {\it empirical plane} of 24\% for the sample of all $78$ clusters
is reduced to a scatter around the dia\-gnostically
selected {\it fundamental plane} of 21 \% for
the sample of 29 clusters with less substructure. Both are
notably amplified when the $L_x$ dependence is ignored.\\
For future work, this scatter can be analyzed in more detail, if one takes
into account different heating mechanisms for the intergalactic gas,
different profiles of the intergalactic gas and the dark matter,
and different populations of galaxies within single clusters.
Furthermore, the scatter around the planes
may be used as an indicator of the evolution
of clusters, if one applies the concept of the {\it fundamental plane}
to samples of clusters belonging to different redshift ranges.

\begin{acknowledgements}

The ROSAT project is supported by the German Bundesministerium
f\"ur Bildung, Wissenschaft, Forsch\-ung und Technologie (BMBF/DARA)
and the Max--Planck--Society.
This research has made use of the NASA/IPAC Extragalactic Data Base
(NED) which is operated by the Jet Propulsion Laboratory, California
Institute of Technology, under contract with the National Aeronautics
and Space Administration. TB is supported by the {\it Sonderforschungsbereich
375 f\"ur Astro--Teilchenphysik der Deutschen Forschungsgemeinschaft} and
thanks Claus Beisbart, Ralf Bender, Stephanie C\^ot\'e and Martin Kerscher
for useful discussions.
Thanks also to Alberto Cappi and Sophie Maurogordato for providing their data,
and in particular to Hans B\"ohringer for his suggestions and constant interest.
Finally, the referee Christophe Adami is acknowledged for many constructive
comments.

\end{acknowledgements}

\newpage

\begin{figure*}
\centerline{\epsfig{figure=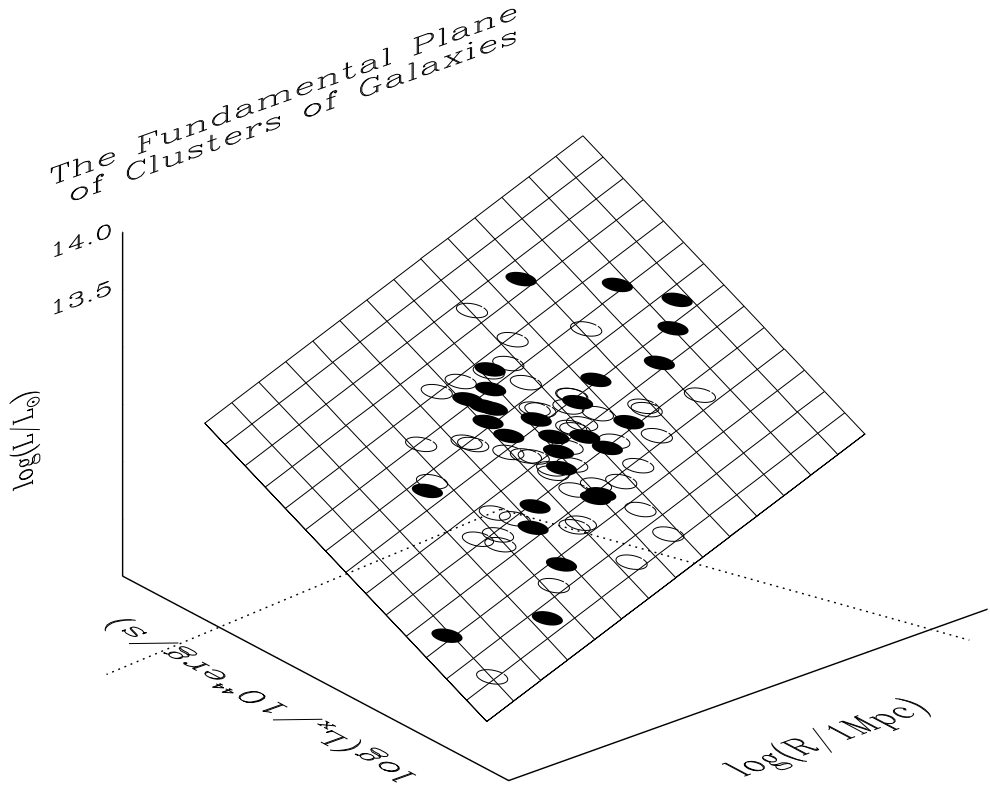,width=12cm,angle=90}}
\caption{
The {\it empirical plane} for clusters of galaxies fitted to the whole sample in
{\it face--on} projection
(shown in the $\log(L_{0}/L_{\odot}) - \log(L_{x}/10^{44} {\rm erg}/{\rm s}) -
\log(R_{0})$--space).
The grid--plane results from a 2--dimensional fit to the data.
The black symbols mark clusters of galaxies with less substructure in the
optical and X--ray energy bands (members of the {\it fundamental plane}).
The white symbols mark the complementary sample of clusters with more
substructure.}

\label{fplane_3dim}

\end{figure*}

\begin{figure*}
\centerline{\epsfig{figure=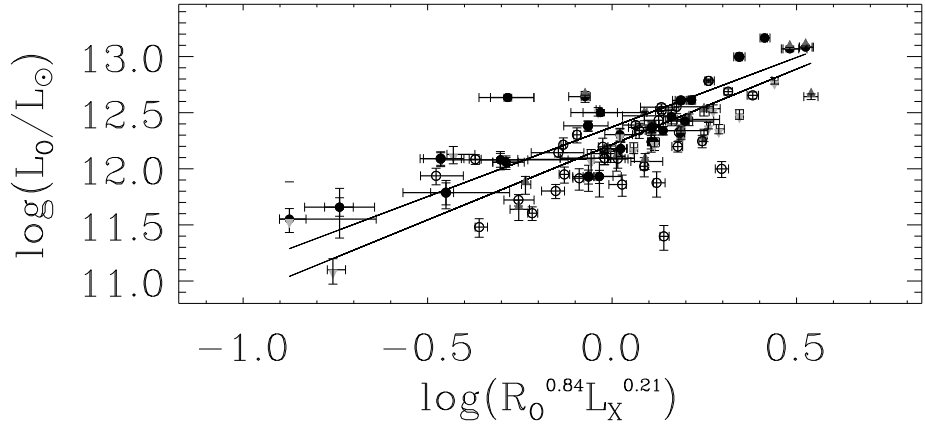,width=7.0cm,angle=90}}
\caption{
The {\it empirical plane} ($L_0 = R_{0}^{0.84} \; L_x^{0.21}$)
for clusters of galaxies fitted to the whole sample in {\it edge--on}
projection (marked by the line with steeper slope).
Again, the black dots mark the clusters with less substructure,
whereas open circles mark clusters with more substructure in both the
optical and X--ray energy bands.
The line with shallower slope represents the fit to the more ``relaxed''
clusters defining the {\it fundamental plane}}.

\label{edge_on}

\end{figure*}

\newpage

\begin{figure*}
\centerline{\epsfig{figure=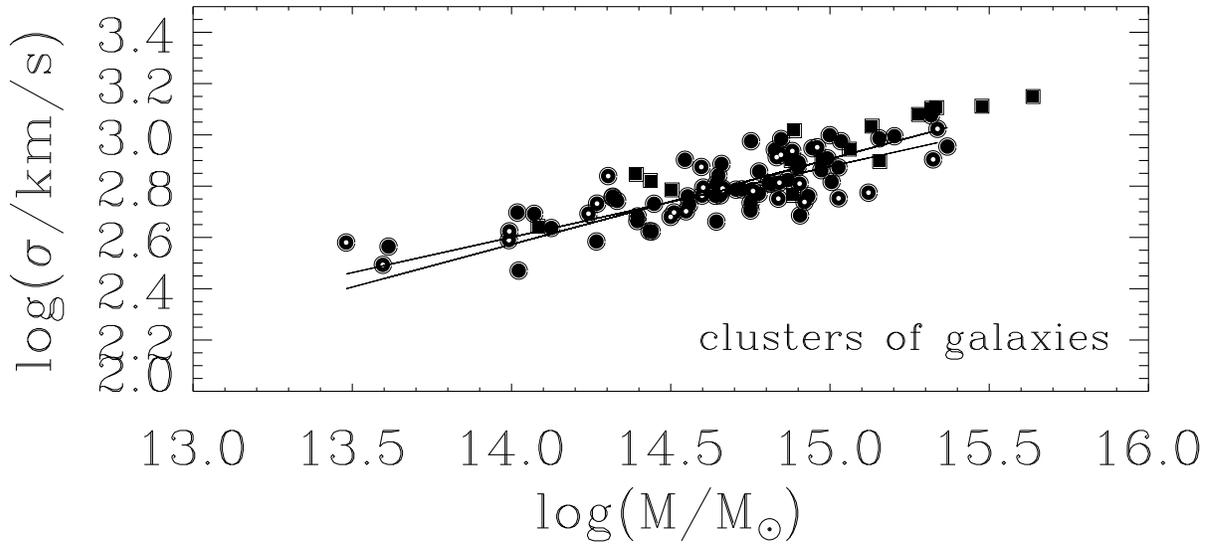,width=9cm,angle=90}}
\caption{
The velocity dispersion versus virial mass.
The velocity dispersion for the clusters is derived from
the measured X--ray luminosities using the empirical relations
between the X--ray luminosity and the temperature
$kT \propto L_{x}^{0.354}$ given by White (1996).
The circles denote our clusters (including a white dot, if belonging to
the {\it fundamental plane}). The squares denote
the clusters with effective radii taken from Cappi \&
Maurogordato (priv. comm.) and velocity dispersion taken from Struble \& Rood (1991).
The line with steeper slope represents the fit to all clusters
corresponding to the {\it empirical plane}.
The other line represents the fit to the more ``relaxed'' clusters
defining the {\it fundamental plane}.}

\label{sigma_m2}

\end{figure*}

\begin{figure*}
\centerline{\epsfig{figure=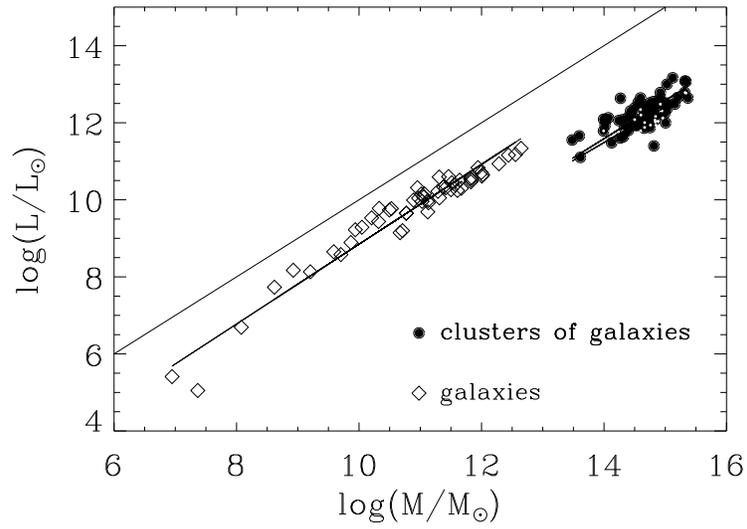,width=8cm,angle=90}}
\caption{
The {\it empirical plane} fitted to the whole sample in
the $\log(L_0/L_{\odot}) - \log(M/M_{\odot}) $--projection
for structures on different scales.
Black circles mark clusters of galaxies and white squares mark
elliptical galaxies taken from Bender et al. (1992).
The upper line corresponds to objects, which would consist
only of optically luminous matter.}

\label{hierclust}

\end{figure*}

\begin{table*}

\caption{
List of the clusters with the luminosities $L_0$ (optical)
and $L_x$ (X--ray), and the optical half--light radii $R_0$.}

\label{sample}

{\normalsize
\setlength{\doublerulesep}{+2pt}
\arrayrulewidth0.5pt
\begin{tabular*}{0.91\hsize}{|l@{\extracolsep\fill}|l|c|c|}
\hline
& & & \\[-7pt]
cluster\hphantom{xxxxx} & $L_0$ & $L_x$ & $R_0$
\\
& $[10^{10} L_{\odot}]$ & $[10^{44}{\rm erg/s}]$ & $[{\rm Mpc}]$
\\
& & & \\[-7pt]
\hline
\hline
& & & \\[-7pt]
   a0076 & $   143.074 $  & $     1.803 $  & $     0.833 $ 
\\
   a0119 & $   451.292 $  & $    11.061 $  & $     1.559 $ 
\\
   a0147 & $   157.714 $  & $     1.462 $  & $     1.482 $ 
\\
   a0160 & $   220.267 $  & $     0.949 $  & $     1.240 $ 
\\
   a0168 & $   210.932 $  & $     2.665 $  & $     1.286 $ 
\\
   a0189 & $    99.819 $  & $     1.340 $  & $     2.098 $ 
\\
   a0195 & $   134.423 $  & $     0.307 $  & $     0.413 $ 
\\
   a0260 & $   290.824 $  & $     1.294 $  & $     1.457 $ 
\\
   a0261 & $    12.599 $  & $     0.067 $  & $     0.247 $ 
\\
   a0295 & $    43.532 $  & $     0.653 $  & $     0.556 $ 
\\
   a0376 & $   233.546 $  & $     4.204 $  & $     0.939 $ 
\\
   a0400 & $    25.010 $  & $     1.233 $  & $     1.394 $ 
\\
   a0407 & $   199.971 $  & $     2.302 $  & $     1.607 $ 
\\
   a0533 & $   124.108 $  & $     0.712 $  & $     1.030 $ 
\\
   a0576 & $   172.687 $  & $     4.679 $  & $     0.932 $ 
\\
   a0779 & $    85.412 $  & $     0.345 $  & $     1.188 $ 
\\
   a0999 & $    82.340 $  & $     0.136 $  & $     1.290 $ 
\\
   a1100 & $   121.335 $  & $     0.688 $  & $     0.940 $ 
\\
   a1139 & $   123.110 $  & $     0.706 $  & $     1.138 $ 
\\
   a1142 & $    72.391 $  & $     0.720 $  & $     1.170 $ 
\\
   a1177 & $    74.972 $  & $     0.513 $  & $     1.645 $ 
\\
   a1185 & $   153.499 $  & $     0.938 $  & $     1.193 $ 
\\
   a1213 & $   151.170 $  & $     0.925 $  & $     1.088 $ 
\\
   a1228 & $   241.229 $  & $     0.266 $  & $     1.165 $ 
\\
   a1314 & $   201.711 $  & $     0.990 $  & $     1.063 $ 
\\
   a1367 & $   157.068 $  & $     3.069 $  & $     0.705 $ 
\\
   a1644 & $   296.538 $  & $     9.702 $  & $     0.720 $ 
\\
   a1656 & $   487.000 $  & $     8.000 $  & $     1.409 $ 
\\
   a1736 & $   322.886 $  & $     8.790 $  & $     1.150 $ 
\\
   a1983 & $   410.034 $  & $     1.237 $  & $     1.713 $ 
\\
   a2052 & $   122.214 $  & $     8.781 $  & $     0.600 $ 
\\
   a2063 & $   439.425 $  & $     6.862 $  & $     2.711 $ 
\\
   a2107 & $   120.115 $  & $     1.755 $  & $     0.379 $ 
\\
   a2147 & $   306.000 $  & $    10.023 $  & $     1.448 $ 
\\
   a2148 & $   270.398 $  & $     0.779 $  & $     1.504 $ 
\\
   a2151 & $   356.908 $  & $     3.300 $  & $     1.198 $ 
\\
   a2152 & $   163.925 $  & $     0.488 $  & $     0.833 $ 
\\
   a2162 & $    61.319 $  & $     0.091 $  & $     0.530 $ 
\\
   a2197 & $   336.479 $  & $     0.463 $  & $     1.093 $ 
\\
   a2199 & $  1171.253 $  & $    15.552 $  & $     1.886 $ 
\\
   a2572 & $   105.208 $  & $     2.168 $  & $     1.047 $ 
\\
   a2589 & $    89.101 $  & $     3.724 $  & $     0.505 $ 
\\
   a2593 & $   176.247 $  & $     3.965 $  & $     1.384 $ 
\\
   a2634 & $   226.430 $  & $     2.614 $  & $     1.738 $ 
\\
   a2657 & $   116.194 $  & $     5.942 $  & $     0.819 $ 
\\
   a2666 & $   200.826 $  & $     0.137 $  & $     1.266 $ 
\\
   a2717 & $  1213.151 $  & $     3.787 $  & $     3.020 $ 
\\
   a2806 & $   209.367 $  & $     0.619 $  & $     1.880 $ 
\\
   a2870 & $   328.938 $  & $     0.357 $  & $     1.870 $ 
\\
   a2877 & $   316.834 $  & $     1.032 $  & $     0.910 $ 
\\
   a3193 & $    45.509 $  & $     0.029 $  & $     0.320 $ 
\\
\hline
\end{tabular*}

}

\end{table*}

\begin{table*}

\label{coeff_page}
 
{\normalsize
\setlength{\doublerulesep}{+2pt}
\arrayrulewidth0.5pt
\begin{tabular*}{0.91\hsize}{|l@{\extracolsep\fill}|l|c|c|}
\hline
& & & \\[-7pt]
cluster\hphantom{xxxxx} & $L_0$ & $L_x$ & $R_0$
\\
& $[10^{10} L_{\odot}]$ & $[10^{44}{\rm erg/s}]$ & $[{\rm Mpc}]$
\\
& & & \\[-7pt]
\hline
\hline
& & & \\[-7pt]
   a3225 & $   123.177 $  & $     0.137 $  & $     0.460 $ 
\\
   a3341 & $    52.744 $  & $     0.677 $  & $     0.550 $ 
\\
   a3367 & $    72.414 $  & $     0.558 $  & $     0.608 $ 
\\
   a3376 & $   218.982 $  & $     5.588 $  & $     0.950 $ 
\\
   a3381 & $   121.501 $  & $     0.022 $  & $     0.940 $ 
\\
   a3389 & $  1464.532 $  & $     0.815 $  & $     3.270 $ 
\\
   a3390 & $   114.425 $  & $     0.309 $  & $     0.610 $ 
\\
   a3395 & $   406.866 $  & $     6.350 $  & $     1.050 $ 
\\
   a3554 & $    86.541 $  & $     0.326 $  & $     0.358 $ 
\\
   a3558 & $   609.462 $  & $    29.090 $  & $     1.440 $ 
\\
   a3560 & $   439.546 $  & $     2.673 $  & $     0.640 $ 
\\
   a3577 & $   136.677 $  & $     1.230 $  & $     0.940 $ 
\\
   a3706 & $   242.537 $  & $     0.693 $  & $     2.242 $ 
\\
   a3716 & $   431.446 $  & $     0.492 $  & $     0.550 $ 
\\
   a3736 & $    85.228 $  & $     0.327 $  & $     1.110 $ 
\\
   a3744 & $   353.983 $  & $     1.265 $  & $     1.360 $ 
\\
   a3747 & $   343.648 $  & $     0.285 $  & $     2.900 $ 
\\
   a3816 & $   174.029 $  & $     0.882 $  & $     1.380 $ 
\\
   a4038 & $   266.279 $  & $     6.633 $  & $     1.070 $ 
\\
   a4049 & $   245.423 $  & $     0.215 $  & $     1.750 $ 
\\
   a4059 & $   275.124 $  & $    11.593 $  & $     0.960 $ 
\\
   a893004 & $    30.240 $  & $     0.159 $  & $     0.590 $ 
\\
   s0141 & $   139.191 $  & $     0.219 $  & $     0.981 $ 
\\
   s0316 & $    40.202 $  & $     0.086 $  & $     1.020 $ 
\\
   s0585 & $   995.286 $  & $     0.628 $  & $     2.890 $ 
\\
   s0639 & $   606.948 $  & $     0.530 $  & $     2.400 $ 
\\
   s0892 & $    63.376 $  & $     0.273 $  & $     0.910 $ 
\\
   s1065 & $    35.631 $  & $     0.082 $  & $     0.170 $ 
\\
\hline
\end{tabular*}
}

\end{table*}

\end{document}